\documentstyle[12pt]  {ioplppt}
\begin{document}
\jl{1}
\letter{On surface properties of two-dimensional percolation clusters}
%[Surface properties of percolation clusters]
\author{S L A de Queiroz\ftnote{1}{E-mail: sldq@portela.if.uff.br}}

\address{Instituto de F\'\ii sica, UFF, Avenida Litor\^anea s/n,
Campus da Praia Vermelha, 24210--340 Niter\'oi RJ, Brazil}

\begin{abstract}
The two-dimensional site percolation problem is studied by transfer-matrix
methods on finite-width strips with free
boundary conditions. The relationship between correlation-length amplitudes and
critical indices, predicted by conformal invariance, allows a very precise
determination of the surface decay-of-correlations exponent,
$\eta_s = 0.6664 \pm 0.0008$, consistent
with the analytical value $\eta_s = 2/3$. It is found that a special
transition does not occur in the case, corroborating earlier series results. At
the ordinary transition, numerical estimates are consistent with the exact
value $y_s = -1$ for the irrelevant exponent.
\end{abstract}

%64.60.Ak Renormalization-group, fractal, and percolation studies of phase
%        transitions
%05.50.+q Lattice theory and statistics; Ising problems
%75.30.Pd Surface magnetism
\maketitle

\nosections
Finite-size scaling concepts are a powerful tool for the determination of
critical properties at phase transitions~\cite{fs1}, especially when coupled
to  phenomenological renormalisation~\cite{fs2} and conformal
invariance~\cite{cardy} ideas. Here we present results from numerical
transfer-matrix calculations of the correlation length, and quantities derived
therefrom, for site percolation on infinite strips with free boundary
conditions (FBC). The use of FBC allows one to assess surface critical
properties, including so-called special and surface transitions~\cite{binder}
when they occur.

Our transfer-matrix formulation of the percolation problem relies on the direct
application of connectivity concepts~\cite{earlyd}, as opposed to taking the
$s \to 1$ limit of the $s$~-state Potts model~\cite{bn}, which corresponds
to bond percolation (and should then, by universality, give the same exponents
as for the
site problem). While the latter approach benefits from being a systematic
expansion in terms of Whitney polynomials, it is devised for general,
continuous
$s$ and thus carries a high degree of inherent complexity. As shown in earlier
work~\cite{earlyd,dds,dst} and below, the geometric picture based on cluster
connectivity allows for a straightforward algorithm to be built, from
which a nicely extrapolating sequence of finite-size estimates is extracted.

We use strips of width $L \leq 10$ sites, both for square and triangular
lattices. This is the same maximum width reached with periodic boundary
conditions (PBC)~\cite{dds,dst}, though in the present case lower symmetry
implies
that the matrices are of considerably larger dimension than on a cylindrical
geometry. First, standard phenomenological renormalisation (PR)~\cite{fs2} is
performed on the site occupation probability $p$ between strips of consecutive
widths, from which estimates of the critical concentration $p_c$, the
temperature-like exponent $y_p$ and the surface decay-of-correlations exponent,
$\eta_s$ are obtained. Alternative finite-size sequences for $\eta_s$
are produced by setting $p$ at the exact (or extrapolated) $p_c$. We then
search for a special transition, by introducing a distinct probability
$p_s$ for site  occupation along the strip boundaries. A two-parameter
PR analysis is carried out, by comparing
correlation lengths on three strips of consecutive widths~\cite{dh,vyj,dqy}.
Only one non-trivial fixed point, shown to correspond to an ordinary
transition, is found upon numerical examination of the recursion
 relations. No evidence is found for the existence of a multicritical point
related to a special, or surface, transition.

The exponent that
governs the decay of correlations along the surface of a semi-infinite plane at
criticality, $\eta_s$, is related to the correlation length on a strip with FBC
by:
\begin{equation}
\eta_s = {2L \over \pi \xi_L(p_c)} \ ,
\label{eq:1}
\end{equation}
a result from conformal invariance~\cite{cardyf1,cardyf2}. Note that for a
triangular lattice with FBC the strip width $L = N \sqrt{3}/2$, where $N$ is
the
number of sites across the strip.

Corrections to scaling must be dealt with, since e.g. \Eref{eq:1} is expected
to be valid only asymptotically. Throughout this work, extrapolations toward
 $L \to \infty$ have been done using the Bulirsch-Stoer (BST)
algorithm~\cite{bst,ms}. As extensively discussed elsewhere~\cite{ms},
whenever the leading correction-to-scaling exponent $\omega$ is not
known {\it a priori} BST extrapolations rely on keeping it as a free parameter
within an interval guessed to be reasonable.
Central estimates and error bars are evaluated
self-consistently by selecting the range of $\omega$ for which overall
fluctuations are minimised. In the following we have allowed
 $0.45 \leq \omega \leq 2.4$ for all quantities, as the most likely
candidates in the case are $\omega = 1$ and 2 (see below).

We implement standard, one-parameter, PR in the usual way by looking for the
fixed point $p^{\ast}$ of the implicit recursion relation:
\begin{equation}
{\xi_L (p^{\ast}) \over L} = {\xi_{L-1} (p^{\ast}) \over L-1} \ ,
\label{eq:2}
\end{equation}
where $\xi_L (p) = -1 / \ln \Lambda_L (p)$ is given in terms of the largest
eigenvalue $\Lambda_L (p)$
of the column-to-column transfer matrix~\cite{earlyd}; $p^{\ast}$ is thus a
finite-size estimate of $p_c$. At the fixed point,
the temperature-like exponent $y_p = 1/\nu$ is evaluated by taking suitable
derivatives~\cite{fs1}. For consistency, $\eta_s$ is obtained from  \Eref{eq:1}
with $\xi$ calculated at $p^{\ast}$.
\begin{table}
\caption{
Results from one-parameter PR.  Uncertainties in last quoted digits are shown
in parentheses. Extrapolations obtained by BST algorithm
with correction-to-scaling exponent $\omega$ in ranges shown. Expected values
are exact, unless otherwise noted. }
\begin{indented}
\item[]\begin{tabular}{@{}lllllll}
\br
&\centre{3}{Square}&\centre{3}{Triangular}\\ \ns
&\crule{3} & \crule{3}\\
$L$ & $p^{\ast}$& $y_p$ & $\eta_s$& $p^{\ast}$& $y_p$ & $\eta_s$\\
\mr
3 & 0.671130 & 0.662822 &  0.290469& 0.573092 & 0.665596 &  0.337283\\
4 & 0.644177 & 0.676427 &  0.351082& 0.547377 & 0.681787 &  0.393936\\
5 & 0.629524 & 0.686306 &  0.394701& 0.533471 & 0.692689 &  0.434325\\
6 & 0.620566 & 0.693798 &  0.427621& 0.525062 & 0.700557 &  0.464391\\
7 & 0.614644 & 0.699685 &  0.453367& 0.519561 & 0.706500 &  0.487600\\
8 & 0.610502 & 0.704439 &  0.474064& 0.515749 & 0.711147 &  0.506047\\
9 & 0.607478 & 0.708362 &  0.491070& 0.512990 & 0.714869 &  0.521047\\
10& 0.605197 & 0.711655 &  0.505297& 0.510928 & 0.717880 &  0.533458\\
Expected   & 0.592745(2)$^a$ & 3/4 & 2/3$^b$ & 1/2 & 3/4 &2/3$^b$ \\
Extrapolated& 0.5925(5) & 0.750(2) & 0.666(3)& 0.5005(2) & 0.750(2) &
0.676(3)\\
$\omega$    &  1.50(50) & 1.10(10) &  1.00(5) & 2.00(10) &  1.10(10) &
1.05(5)\\
\br
\end{tabular}
\item $^{\rm a}$ Monte Carlo \protect{~\cite{ziff}}
\item $^{\rm b}$ Predicted by conformal invariance \protect{~\cite{cardyf2}}
\end{indented}
\end{table}

Our results are shown in table 1, where the values of $p^{\ast}$ and $y_p$
for $L = 3$ and $4$ on the square lattice have been obtained
previously~\cite{earlyd}.  The amplitude of finite-size corrections
is much larger than for the corresponding cases of PBC
(see e.g. Table I of reference~\cite{dst}). However, the finite-lattice
sequences are generally well-behaved, allowing for smooth extrapolations.
Comparing our extrapolated estimates for $p_c$ and $y_p$ with the well-known
respective values provides a good overall check of the reliability of our
procedures. For the square lattice our $p_c$ agrees very well with, but is
less precise than, the best estimate for the percolation threshold
$p_c= 0.592745 \pm 0.000002$~\cite{ziff}. A similar picture holds for the
comparison with the exact
$p_c= 1/2$ for the triangular lattice, and $y_p= 3/4$  (both lattices).

Turning now to $\eta_s$, table 1 provides a direct test of the prediction
$\eta_s = 2/3$~\cite{cardyf2}, previously confirmed only indirectly via the
series result $\gamma_1 = 2.10 \pm 0.02$~\cite{debell}
which, together with the scaling relation $2\gamma_1 = \gamma + \nu(2-\eta_s)$
 and the exact values $\gamma = 43/18$
and $\nu = 4/3$ gives $\eta_s = 0.64 \pm 0.03$ ~\cite{cardyf2} . Though the
agreement is generally very satisfactory, the sequence for the triangular
lattice seems to extrapolate towards a region slightly above 2/3.

 In order to improve the quality of our estimates~\cite{dds,dqrbs},
we have also generated sequences of finite-size data for $\eta_s$ by
setting $p$ at the best available (or exact) value of $p_c$. For
comparison with the corresponding data
of reference~\cite{dst} for the bulk exponent $\eta$,
given by $\eta = L/\pi \xi_L(p_c)$~\cite{cardyf1}
(where $\xi_L$ is related to the largest eigenvalue of the transfer matrix
with PBC and, in reference~\cite{dst}, is calculated at the respective
 $p^{\ast}$ as in table 1 above), we have done the same for
PBC. The results are displayed in table 2.
\begin{table}
\caption{
Results for $\eta_s$ and $\eta$ obtained by setting $p = 0.592745$ (square)
or $p=1/2$ (triangular lattice).  Uncertainties in last quoted digits are
shown in parentheses. Extrapolations obtained by BST algorithm
with correction-to-scaling exponent $\omega$ in ranges shown. Expected
values are exact, unless otherwise noted. }
\begin{indented}
\item[]\begin{tabular}{@{}lllll}
\br
&\centre{2}{Square}&\centre{2}{Triangular}\\ \ns
&\crule{2} & \crule{2}\\
$L$ & $\eta_s$& $\eta$ & $\eta_s$& $\eta$\\
\mr
2 &  0.4326951882 & 0.2163475941 & 0.4673843349 & 0.2114765825\\
3 &  0.4848101732 & 0.2130595008 & 0.5135927770 & 0.2111933048\\
4 &  0.5174925026 & 0.2125576128 & 0.5423526787 & 0.2103549509\\
5 &  0.5400291833 & 0.2114673276 & 0.5619831315 & 0.2097409643\\
6 &  0.5565576800 & 0.2107370714 & 0.5762429521 & 0.2093492678\\
7 &  0.5692194471 & 0.2102232886 & 0.5870747896 & 0.2090956127\\
8 &  0.5792399531 & 0.2098564768 & 0.5955841200 & 0.2089245599\\
9 &  0.5873733739 & 0.2095868033 & 0.6024472809 & 0.2088045137\\
10 & 0.5941102466 & 0.2093833099 & 0.6081026832 & 0.2087173009\\
Expected   & 2/3$^a$ & 5/24 & 2/3$^a$ & 5/24 \\
Extrapolated& 0.6664(4) & 0.20835(2) & 0.665(1)& 0.20833(2) \\
$\omega$    &  1.00(5) & 1.90(10) &  1.00(10) & 2.00(5) \\
\br
\end{tabular}
\item $^{\rm a}$ Predicted by conformal invariance\protect{~\cite{cardyf2}}
\end{indented}
\end{table}

For the square lattice the
central estimate of $p_c$ from reference~\cite{ziff} has been used. Had the
respective error bars been taken into account, this would typically give rise
 to uncertainties in the sixth decimal place of $\eta_s$ or $\eta$. BST
extrapolations of the truncated values point essentially towards the same
 limits exhibited in table 2 (though with error bars roughly doubled),
which shows that the corresponding sequences are rather robust.

The amplitude of finite-size corrections is much larger for FBC than for
PBC, a trend already noticed in the discussion of table 1. Comparison with
the final estimate
$\eta = 0.2088 \pm 0.0008$ of reference~\cite{dst} suggests that, for PBC at
least, setting $p = p_c$ rather than at the approximate $p^{\ast}$ reduces
error bars by one order of magnitude. This latter statement assumes that both
 BST and the extrapolation procedures described in reference~\cite{dst} are of
 comparable intrinsic accuracy, which is reasonable for long ($\simeq 10$
elements) sequences  such as those encountered here (for shorter sequences, BST
 would be relatively more reliable than other methods~\cite{ms}). Turning to
FBC,
one sees a similar
improvement in the extrapolated values of $\eta_s$ relative to those in table
1.
Focusing on the remarkably smooth sequence for the square lattice, and taking
into
account the effect of the uncertainty in $p_c$ on the estimates of $\eta_s$
 as discussed above, we reach the final estimate $\eta_s = 0.6664 \pm 0.0008$.
This is entirely in agreement with the prediction $\eta_s = 2/3$ from conformal
invariance~\cite{cardyf2}, and 1$1\over 2$ orders of magnitude more accurate
than
previous numerical results~\cite{cardyf2,debell}.

It is known from the exact solution of the Ising model that finite-size
estimates of the critical temperature and exponents converge
as $T_c(L) - T_c \sim L^{-3}$; $y(L) -y \sim L^{-2}$ ($y = \nu, \eta$)
for PBC~\cite{dds}, while for FBC the corrections are respectively
proportional to $L^{-2}$ and $L^{-1}$~\cite{burk}. For percolation on strips
with PBC, numerical evidence is similarly consistent with
 $p_c(L) - p_c \sim L^{-3}$ and $y_p(L) - y_p \sim L^{-2}$~\cite{dds}. In the
present case, the data of tables 1 and 2 point towards the following
scenario: $y_p(L) - y_p \sim L^{-1}$, $\eta_s(L) - \eta_s \sim L^{-1}$ (FBC);
$\eta(L) - \eta \sim L^{-2}$ (PBC). Though data for the triangular
lattice indicate $p_c(L) - p_c \sim L^{-2}$ as expected, for the square lattice
one seems to get fits with the same quality either for $\omega = 1$ or 2, or
just about any value in between. We have been unable to sort out this
apparently
discrepant behaviour.

We have investigated the possible existence of a higher-order critical
point, related to a surface-assisted transition. Series work indicates
that a special transition should not be
expected for percolation clusters in two dimensions (though in
three-dimensional
systems it should occur)~\cite{debell,dbl}. The work described below is a
direct
test of such results for the two-dimensional case.

Similarly e.g. to studies of
polymer adsorption~\cite{vyj,dqy}, a distinct occupation probability $p_s$
is introduced for sites on either strip boundary. Fixed points
$(p^{\ast},p_s^{\ast})$ are obtained by comparing correlation lengths on
three strips~\cite{dh}:
\begin{equation}
{\xi_L (p^{\ast},p_s^{\ast}) \over L} = {\xi_{L-1} (p^{\ast},p_s^{\ast}) \over
L-1}
 = {\xi_{L-2} (p^{\ast},p_s^{\ast}) \over L-2} \ .
\label{eq:3}
\end{equation}
By analogy with polymer adsorption, if a special transition occurs it must be
at some $p_s^{\ast} > p^{\ast}$ so that the critical cluster is located
predominantly close to the edge. As $p^{\ast}$ is a bulk quantity, one expects
it to converge to $p_c$ regardless of whether the transition is ordinary or
special.
 By scanning the $(p,p_s)$ space we have ascertained that there is only one
non-trivial solution of \Eref{eq:3}, which
corresponds to an ordinary transition. This can be seen from the estimates
of critical parameters and respective exponents shown in table 3.

\begin{table}
\caption{
Results from two-parameter PR.  Uncertainties in last quoted digits are shown
in parentheses. Extrapolations obtained by BST algorithm
with correction-to-scaling exponent $\omega$ in ranges shown. Expected values
are exact, unless otherwise noted.}
\begin{indented}
\item[]\begin{tabular}{@{}llllll}
\br
\ms
\centre{6}{ (a) Square}\\ %\ns
$L$ & $p^{\ast}$& $p_s^{\ast}$& $y_p$ & $y_s$ & $\eta_s$\\
\mr
5 &  0.595339 &  0.503680 &  0.731343 &  -- 1.03326 &  0.634489\\
6 &  0.595215 &  0.503230 &  0.736897 &  -- 1.04906 &  0.635388\\
7 &  0.594602 &  0.500260 &  0.740391 &  -- 1.04301 &  0.640568\\
8 &  0.594148 &  0.497414 &  0.742624 &  -- 1.03692 &  0.644923\\
9 &  0.593824 &  0.494839 &  0.744160 &  -- 1.03161 &  0.648407\\
10 & 0.593590 &  0.492552 &  0.745265 &  -- 1.02719 &  0.651168\\
Expected   & 0.592745(2)$^a$ & --- & 3/4 & -- 1 &2/3$^b$ \\
Extrapolated & 0.5926(1) & 0.460(2) &  0.750(2) & -- 1.001(1) & 0.666(1)\\
$\omega$ &      2.0(4)   & 0.85(15) &  2.0(2)   & 1.9(1)  &  2.00(5)\\
\br
\ms
\centre{6}{ (b) Triangular}\\ %\ns
$L$ & $p^{\ast}$& $p_s^{\ast}$& $y_p$ & $y_s$ & $\eta_s$\\
\mr
5 & 0.503487 & 0.425365 & 0.734458 & -- 1.022979 & 0.844465\\
6 & 0.502259 & 0.421150 & 0.741249 & -- 1.027182 & 0.802242\\
7 & 0.501519 & 0.417782 & 0.744379 & -- 1.024663 & 0.777288\\
8 & 0.501064 & 0.415120 & 0.746081 & -- 1.021105 & 0.760540\\
9 & 0.500776 & 0.413005 & 0.747086 & -- 1.017650 & 0.748374\\
10& 0.500595 & 0.411374 & 0.747618 & -- 1.013932 & 0.738967\\
Expected   & 1/2 & --- & 3/4 & -- 1 &2/3$^b$ \\
Extrapolated & 0.497(2) & 0.402(1) &  0.750(2) & -- 1.004(3) & 0.680(15)\\
$\omega$ &      2.0(4)   & 2.0(1) &  2.0(4)   & 2.0(4)  &  2.0(4)\\
\br
\end{tabular}
\item $^{\rm a}$ Monte Carlo \protect{~\cite{ziff}}
\item $^{\rm b}$ Predicted by conformal invariance\protect{~\cite{cardyf2}}
\end{indented}
\end{table}
Once again, the smooth convergence of the sequences of estimates of $p^{\ast}$,
$y_p$ and $\eta_s$ towards the expected values confirms that our procedures
are, on the whole, reliable. That the second exponent, $y_s$, is negative
ensures that we are dealing with an ordinary critical point; our
extrapolation is compatible with $y_s = -1$, a result derived on
general grounds for the ordinary transition of two-dimensional
systems~\cite{bc}. The
non-universal $p_s^{\ast}$ converges to values smaller than the
respective $p_c$ for each lattice. This resembles the ordinary transition for
polymers, at which the fugacity for surface contacts is slightly
{\it de-}enhanced~\cite{vyj,dqy} .

With the notable exception of the sequence for $p_s^{\ast}$ for the square
lattice, the leading correction-to-scaling exponent seems to be in the
vicinity of 2, or even larger, for all quantities involved. At present
it is not clear whether this
feature is fortuitous, or in some way related to the structure of the
two-parameter PR equations.

We have shown that the exponent that controls the decay of critical
correlations
along the surface of a semi-infinite percolating plane is
$\eta_s = 0.6664 \pm 0.0008$, consistent with the prediction from conformal
invariance $\eta_s = 2/3$. By setting the site occupation probability $p$
at its critical value $p_c$, clean numerical evidence has been provided that
the
finite-size estimates of $\eta_s$ and of the bulk exponent $\eta$ scale
respectively as
$\eta_s(L) - \eta_s \sim L^{-1}$ (FBC); $\eta(L) - \eta \sim L^{-2}$ (PBC).
It has been
shown by numerical examination of suitable two-parameter PR recursion relations
 that no
special transition occurs in the case; further, at the ordinary critical point
the irrelevant exponent is, with all probability, $y_s = -1$ exactly.

Extensions of the present work to branched polymers (lattice
animals)~\cite{dds}
 are currently being pursued. Though conformal invariance concepts are not
applicable in the case~\cite{mdb}, surface critical indices such as the
 crossover exponent $\phi = y_s/y$ can be calculated and compared e.g. to
series results~\cite{dpffs}, for which error bars are rather large at present.

\ack
The author thanks M Henkel and J L Cardy for interesting conversations, and
Departamento de F\'\ii sica, PUC/RJ for use of their computational facilities.
This research is supported by CNPq, FINEP and CAPES.

\end{document}